\documentclass[a4paper]{jpconf}
\usepackage{graphicx}
\begin{document}
\title{Turbulence Time Series Data Hole Filling using Karhunen-Lo\`eve and ARIMA methods}

\author{M~P~J~L Chang$^1$, H Nazari$^2$, C~O Font, G~C Gilbreath and E Oh$^3$}

\address{$^1$ Physics Department, University of Puerto Rico, Mayag\"uez, Puerto Rico 00680 \\
	 $^2$ 234 Calle Bellas Lomas, Mayag\"uez, Puerto Rico, 00682 \\
	 $^3$ U.S. Naval Research Laboratory, Washington D.C. 20375
}

\ead{mark@charma.uprm.edu}

\begin{abstract}
Measurements of optical turbulence time series data using unattended instruments over long time intervals inevitably lead to data drop-outs or degraded signals.  We present a comparison of methods using both Principal Component Analysis, which is also known as the Karhunen--Lo\`eve decomposition, and ARIMA that seek to correct for these event-induced and mechanically-induced signal drop-outs and degradations. We report on the quality of the correction by examining the Intrinsic Mode Functions generated by Empirical Mode Decomposition.  The data studied are optical turbulence parameter time series from a commercial long path length optical anemometer/scintillometer, measured over several hundred metres in outdoor environments.
\end{abstract}

\section{Introduction}
How many data points can be missed off (set to zero) from a 1-D discrete, regularly spaced time series record and still be recoverable?  Such a question is prompted by many situations, often associated to the automated collection of data.  We are motivated to address this question by our experimental field work to record the behaviour of the strength of optical turbulence parameter, $C_n^2$, over time intervals of several weeks, under different climate conditions \cite{Santiago05,Font0605050,Chang0701022,Chang06AppOp,Font06MS}.

The data of interest to us are path integrated measures of $C_n^2$ measured across the visible to near infrared region.  The measurement path is a 600 m stretch of free space, approximately 1.5 m altitude above sea water in the Caribbean.  The probe beam originated from an LED centered on 0.9 $\mu$m; the instrumentation are two identical Optical Scientific Inc. LOA-004\footnote{see http://www.opticalscientific.com or email info@opticalscientific.com.} systems, which employ aperture averaging to estimate the value of $C_n^2$.  The LOA-004s are generally left unattended over the course of up to a few days during field operation, which mean that they are susceptible to spurious events which result in gaps in the data record.  Although for some types of data analysis techniques, data gaps of a limited size can be tolerated, this is not universal.  Moreover, there exists a new class of very powerful techniques based on Empirical Mode Decomposition \cite{Huang98,Chang06AppOp} that are exceedingly sensitive to lossy datasets.  It is for this reason we are exploring different methodologies to synthetically fill the data holes; we describe the results from our studies using Principal Component Analysis in this paper.

\section{Karhunen--Lo\`eve or Principal Component Analysis}
\label{sect:PCA}

The principal components of any ensemble can be used to identify the members of that ensemble.  This idea forms the foundation of face recognition and tracking through {\em{eigenfaces}} (see, for example, Turk and Pentland 1991 \cite{Turk91}).  We may extend this method to reconstruct the missing data for any data record under given restrictions.  The key point is that the gappy data record must have the same, or similar, salient features as all the members of the ensemble. 

The principal components are the eigenvectors of the covariance matrix of the data and represent the features of the dataset.  Provided that a reference library can be created, each member of the library will contribute to each eigenvector, more or less.  As such, each member can be exactly represented by a linear combination of eigenvectors.  Any similar data record external to the library will also be represented by a linear combination of eigenvectors, within a margin of error.

The first step for the filling procedure must therefore be to define the reference library.  We may do so by collecting a family of turbulence data series that share certain specific characteristics; a difficult task since the definition of such is an open question.  Moreover, since the mean value of the family of reference data plays a key part, all the members of the library would require normalisation.  Again how to do this is an open question.  Alternatively we may use the neighbouring record around a data hole, sectioning this information to provide the ensemble members.  We opt for the latter technique since the record pre and post the data hole (within a certain time interval) ought to be similar in nature to the missing data.  The mean value of this type of library would probably not differ greatly from the mean of the missing data, so normalisation would not be so crucial.

How does one determine the principal components of a reference library? Following Sirovich and Kirby \cite{Sirovich87}, let the $M$ members (each of length $N$) of the reference ensemble be $\{ \varphi_n \}$.  Thus, the average data record of this ensemble will be 
\begin{equation}
\overline{\varphi} = <\varphi> = \frac{1}{M} \sum_{n=1}^M \varphi_n
\end{equation}
It is very reasonable to assume that departures from the mean record will provide an efficient procedure for extracting the primary features of the data.  Therefore, we define
\begin{equation}
\phi_n = \varphi_n - \overline{\varphi}
\end{equation}
Now, if we consider the dyadic matrix
\begin{equation}
C = \sum_{n=1}^{M} \phi_n \phi^T_n = A A^T 
\end{equation}
where each term of the sum signifies a second rank tensor product, we can recognize this as the ensemble average of the two point correlation of the deviations from the mean. Here, $A^T$ is the transpose of $A$.

We require eigenvectors $u_n$ of the matrix $A A^T$.  For ensembles whose members have a large number of points $N>M$, matrix $A A^T$ issingular and its order cannot exceed $M$.  To find those eigenvectors of $A A^T$ corresponding to nonzero eigen values, Turk and Pentland used a standard singular value decomposition technique, as described below.
\begin{eqnarray}
A^T A v_n & = & \mu v_n \\ \nonumber
A A^T A v_n & = & \mu_n A v_n \\ \nonumber
C A v_n & = & \mu A v_n 
\end{eqnarray}
where $\mu_n$ are the eigenvalues. This deduction can be equated to
\begin{equation}
C u_n = \mu_n u_n
\end{equation}
where $u_n = A v_n$.  Thus $AA^T$ and $A^TA$ have the same eigenvalues and their eigenvectors are related through $u_n = A v_n$, provided that $||u_n|| = 1$.
The treatment described is recognizable as the Karhunen--Lo{\'e}ve (KL) method \cite{Goodman85}.

The implication is that a dataset $\phi^\prime$ can be obtained from a limited summation
\begin{equation}
\phi^\prime \approx \sum_{n=1}^M a_n u_n
\end{equation}
where the coefficients $a_n$ are obtained through the inner product
\begin{equation}
a_n = (\phi^\prime, u_n)
\end{equation}
We emphasise that $\phi^\prime$ is not considered to be part of the $\{ \phi_n \}$, although it is similar in features.

\section{Proof of Concept}
\label{sect:proof}
To demonstrate the validity of the assertion of the previous section, we take a perfect data record of $C_n^2$ measurements over a 7 hour period starting from midnight, smoothed by a forward moving rolling average of 5 minutes' interval (60 data points).  The data contain 2492 points in total.
\begin{figure}[!htp]
\begin{center}
\includegraphics[width=18pc]{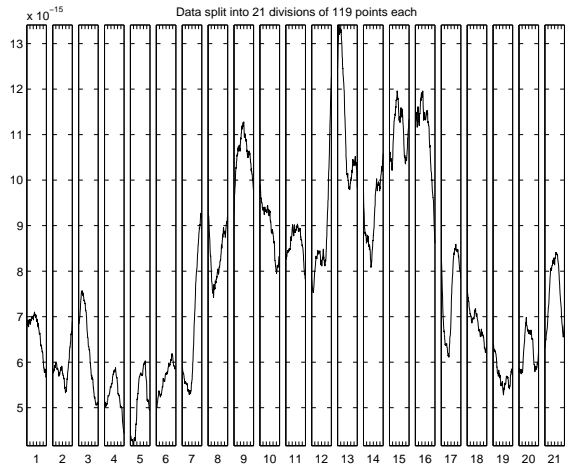}
\caption{\label{fig:Cn2data} The test data, split into 21 sections.  The data are padded before division with a set of points taken from the tail of the time series and mirrored outward.}
\end{center}
\end{figure}
We split up the test data into 21 sections, where all but 1 are members of the reference library, as shown in Fig. \ref{fig:Cn2data}.  The exclusive section is set to zero, and the algorithm described in Sec. \ref{sect:PCA} is employed on the 20 library elements.  The reconstructions are created from the KL coeffiecients of the eigenvectors equivalent to the sections adjacent to the missing data.  Hence we will talk of a prior and posterior reconstruction meaning e.g. for omitted section 5, we use for the prior the KL coefficient equivalent to section 4 and for the posterior reconstruction we use the coefficient equivalent to section 6 of the test data set.  Note that this does not reconstruct those sections, since the eigenvectors are generated from the entire reference library.

\begin{figure}[!htp]
\begin{center}
(a) \includegraphics[width=24pc]{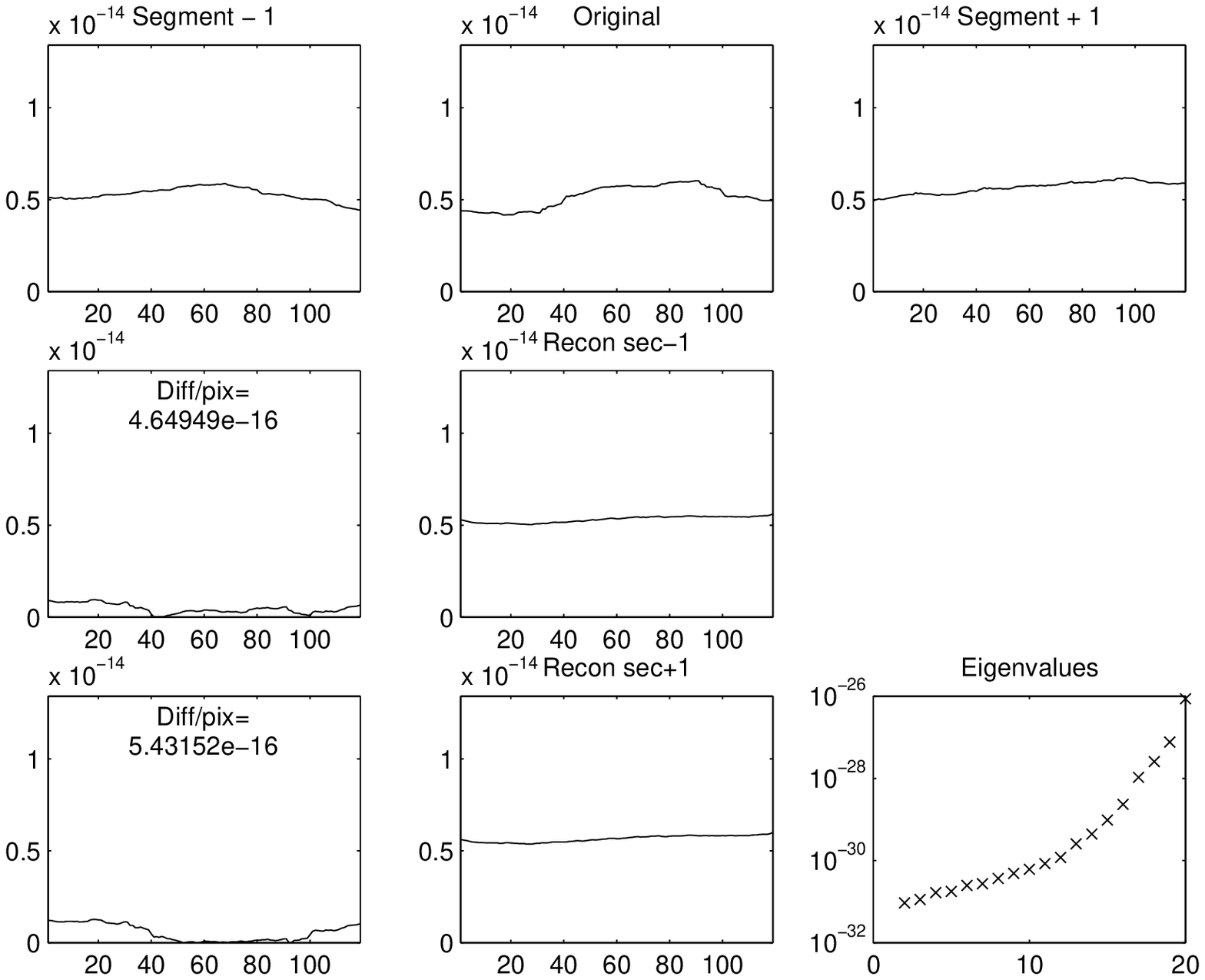} \vspace{2pc} \\
(b) \includegraphics[width=24pc]{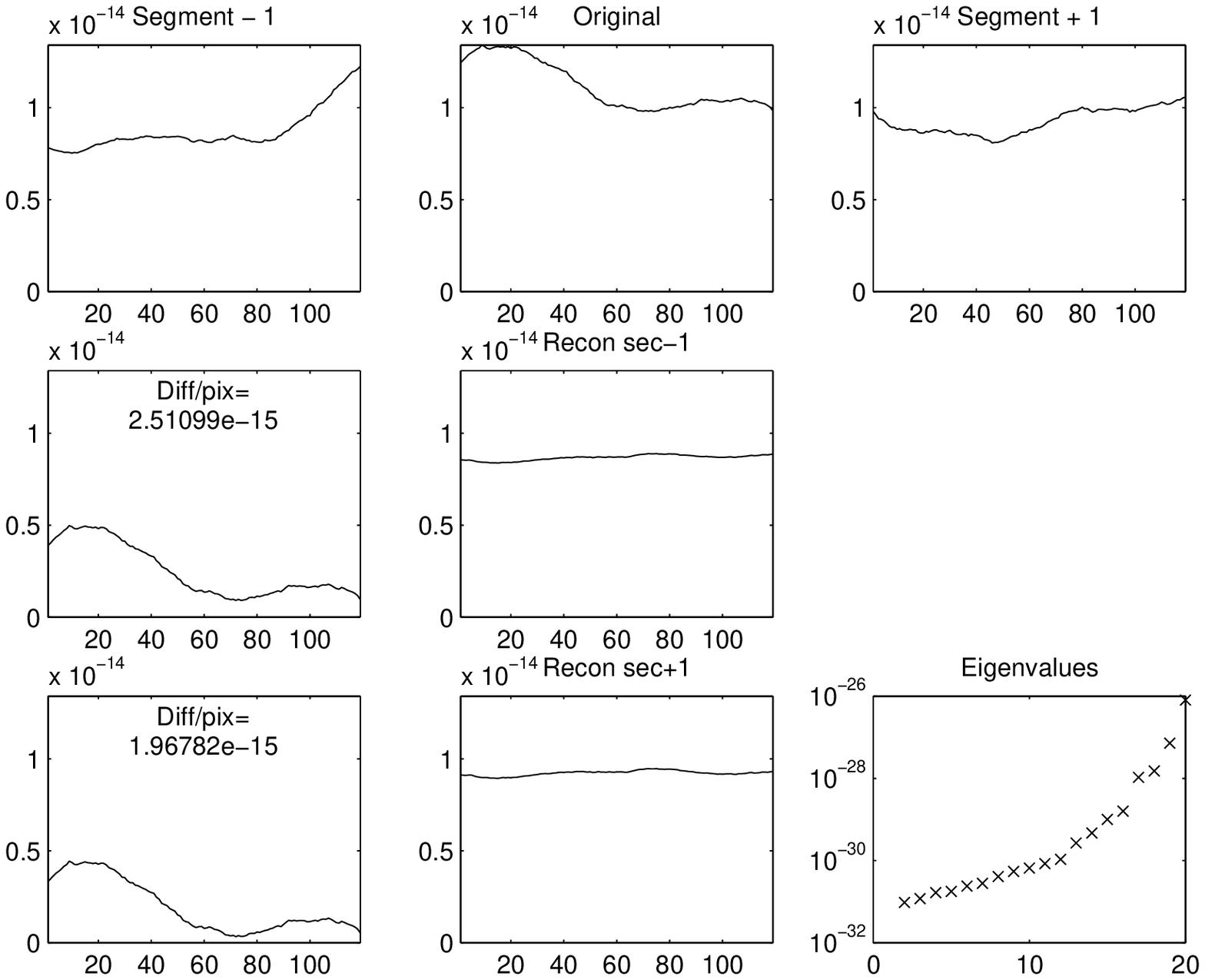}
\caption{\label{fig:KLresult} Best and worst reconstruction result using PCA for (a) section 5 and (b) section 13 respectively. The layout in each set is : (Top Left) Segment prior to the selection.  (Top Middle) The original selected data.  (Top Right) Segment after selection.  (Centre Left) The difference between the reconstruction using the coefficient of eigenvector related to the prior segment and the original.  The value shown is the mean absolute difference per pixel. (Centre Right) The (prior) reconstruction.  (Bottom Left) The difference between the reconstruction using the coefficient related to the posterior segment and the original.  (Bottom Middle) The (posterior) reconstruction.  (Bottom Right) The eigenvalue spectrum.}
\end{center}
\end{figure}

The reconstructions shown in the left hand set of Fig. \ref{fig:KLresult} represent the best level of error, while the right hand set shows the worst.  It is clear that the greater the difference between the (masked off) original data section and its neighbours, the poorer the reconstruction will be.  Nevertheless, the reconstruction errors for the full set of test data are all 1 order of magnitude less than the mean value of the reconstruction and the original data segment.
Evidently the end points have not been synthesised to be continuous with the adjacent segments of the reference library signal, as can be seen from the error.  A continuity condition in terms of the both the function and its derivative has to be imposed on both ends of the reconstructed segment in order to achieve smoothness.  The most effective way to do so is currently being investigated.
The eigenvalues determined through this method show a similar distribution in both cases, with minor variations only at the upper end of the spectrum, implying that the KL differences between best and worst section for reconstruction is not very large.

\subsection{KL and EMD}
In the absence of a workable continuity condition, we present here the effects of crudely patching the data hole with reconstructions determined from the prior and posterior terms with respect to the gap.

\begin{figure}[!htp]
\begin{center}
\begin{tabular}{cc}
\includegraphics[height=0.4\textwidth]{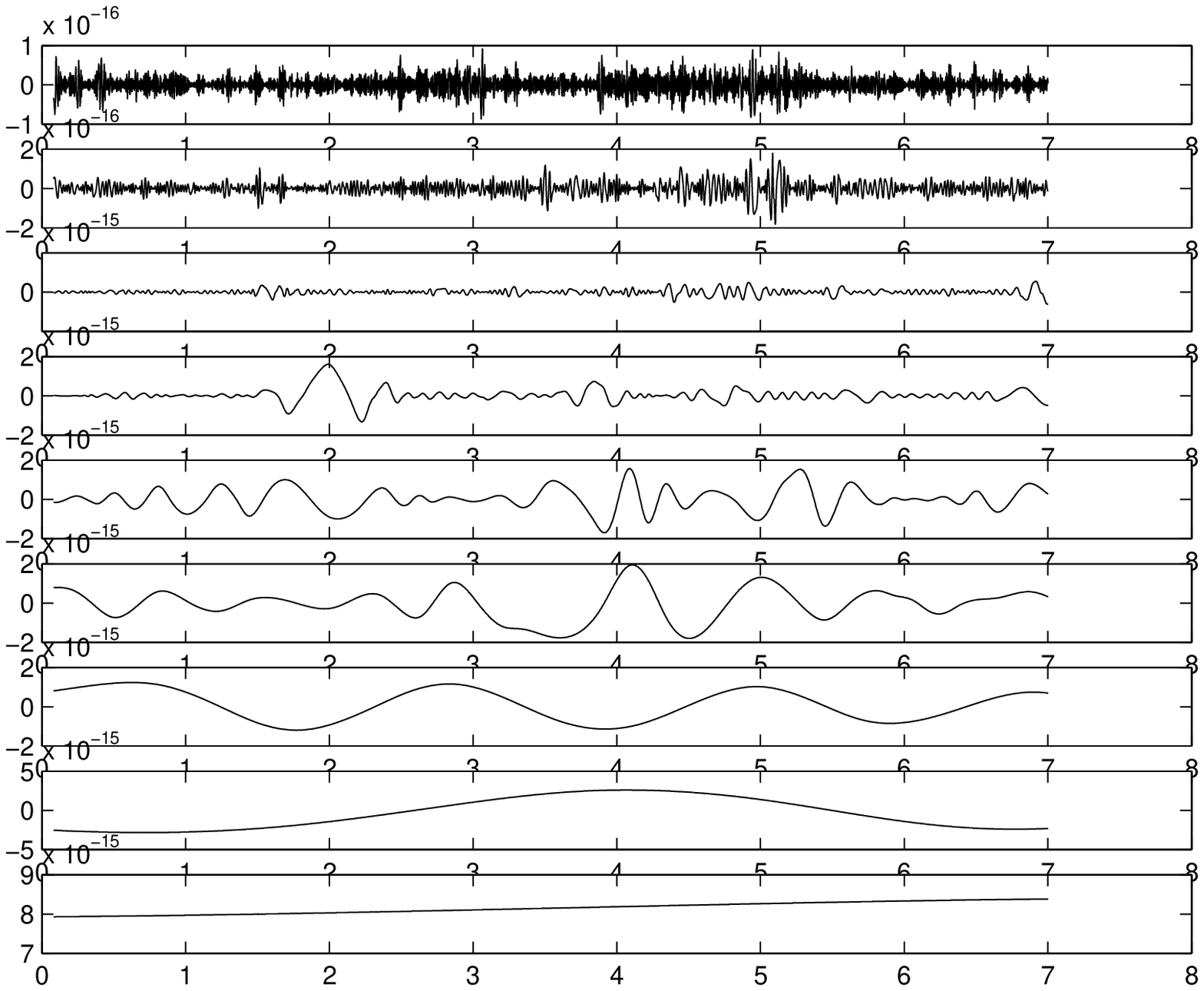} & \includegraphics[height=0.4\textwidth]{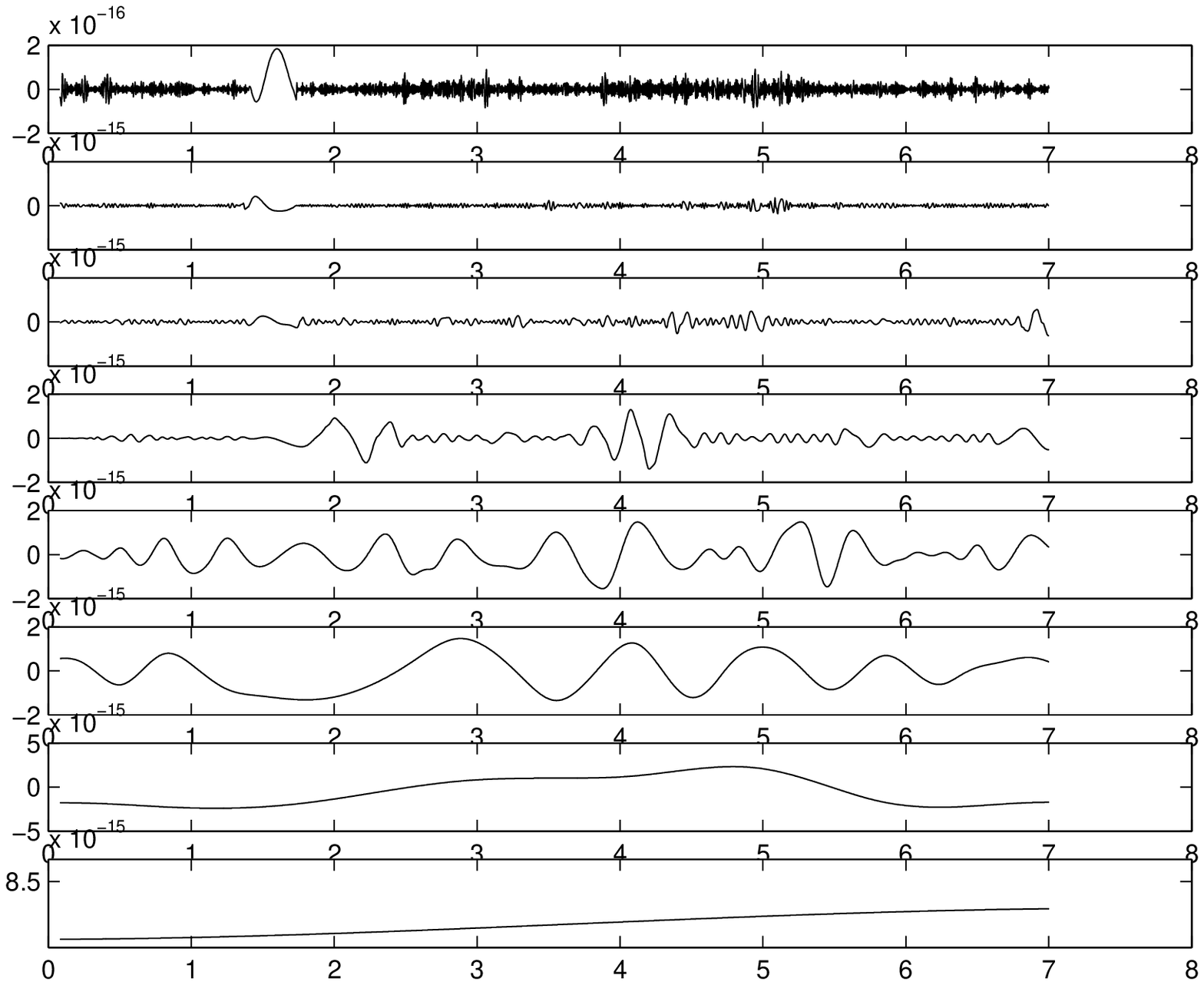} \\
$EMD_O$ & $EMD_{Linear}$ \\ 
\includegraphics[height=0.4\textwidth]{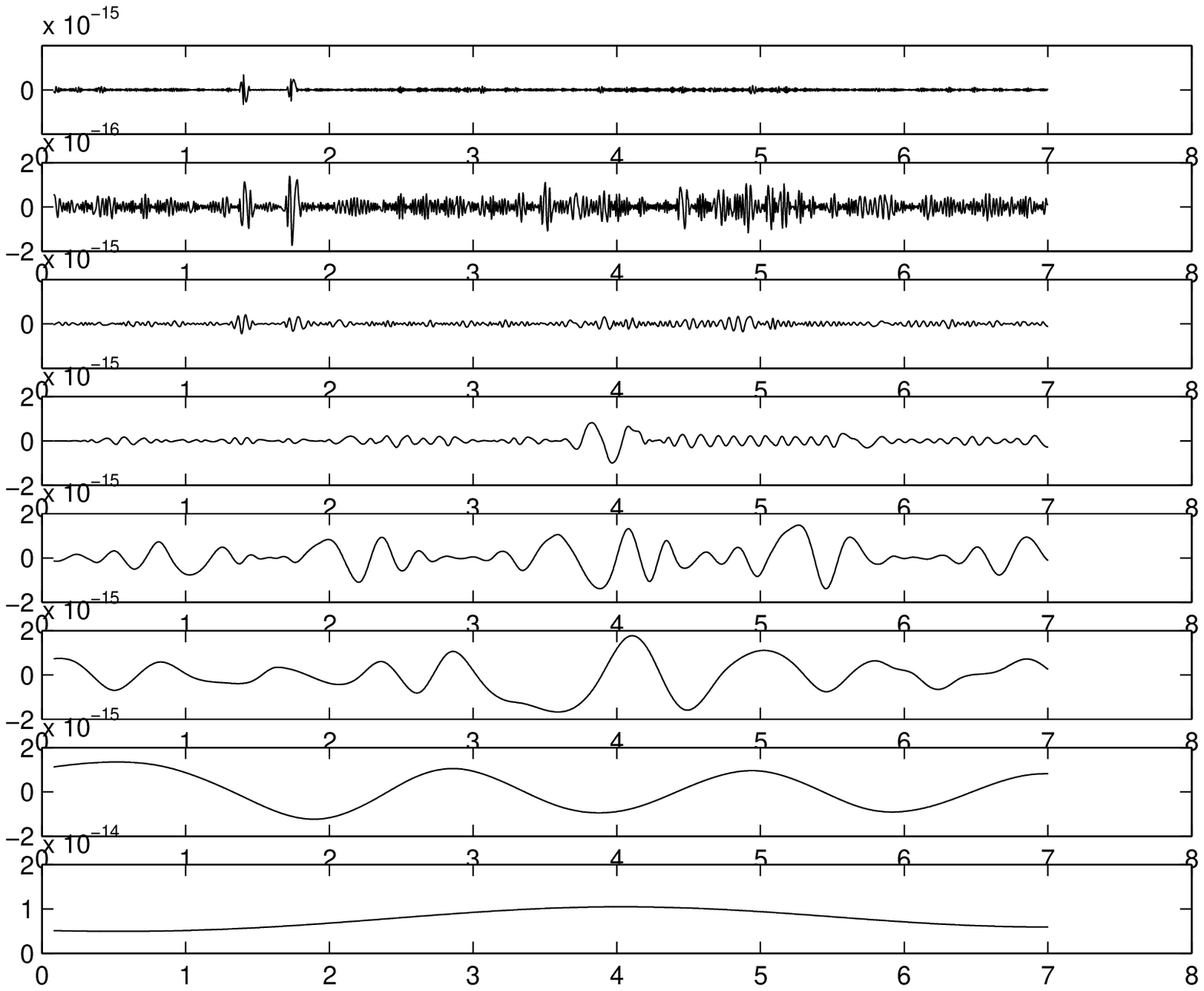} & \includegraphics[height=0.4\textwidth]{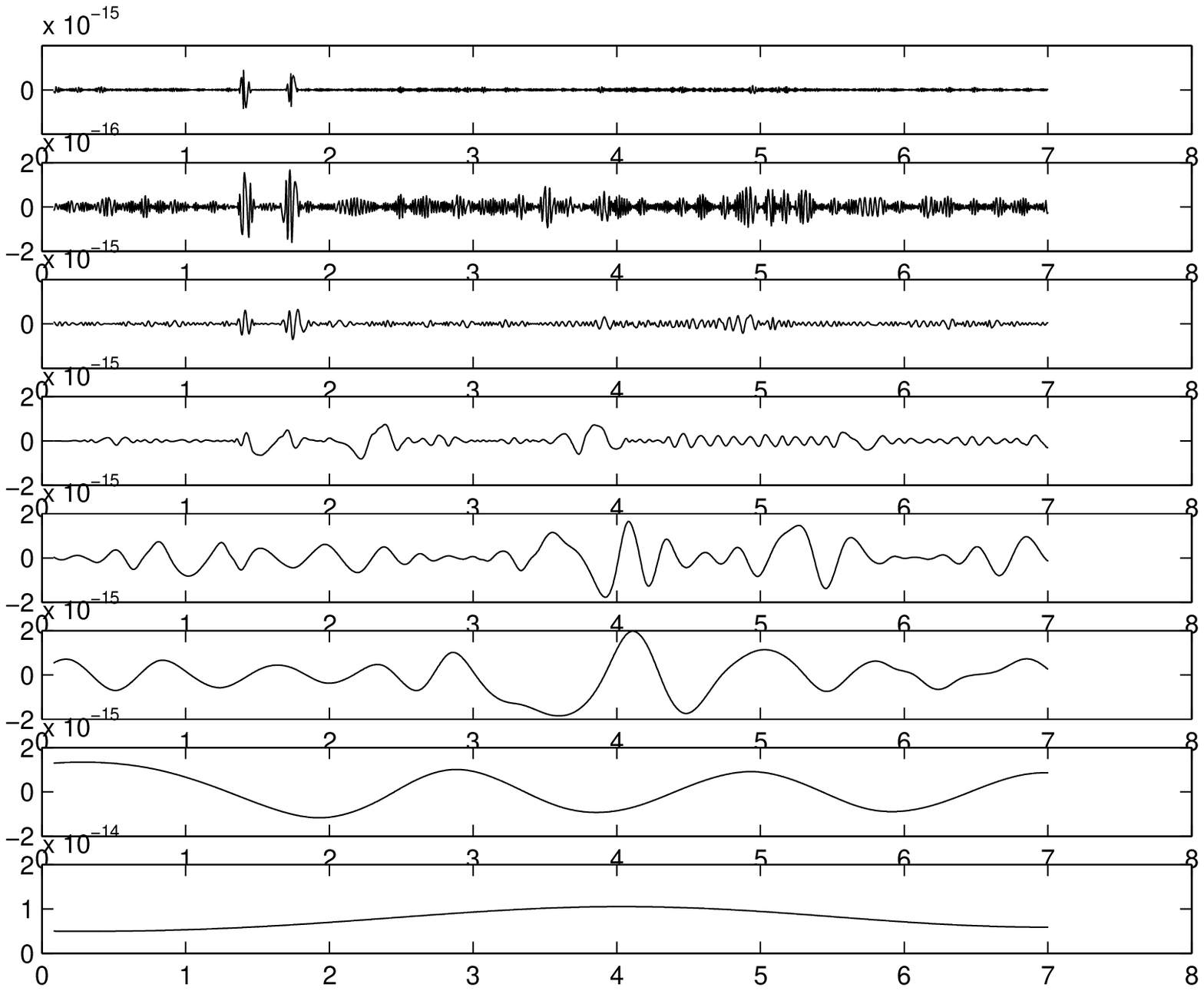} \\
$EMD_L$ & $EMD_R$
\end{tabular}
\end{center}
\caption{\label{fig:EMD} (Top Left) The IMFs 1 to 8 (top to bottom) and the residual trend line of $EMD_O$, generated by applying Empirical Mode Decomposition to the test data.  (Top Right) The IMFs 1 to 7 and the residue of $EMD_L$, generated from data using the PCA reconstruction method with the coefficient of the prior section.  (Bottom Left) The IMFs 1 to 7 and the residue of $EMD_R$, generated from data using the PCA reconstruction method of the posterior section.  (Bottom Right) The IMFs 1 to 7 (top to bottom) and the residue, generated from data with a linear interpolant across section 5 of the test data.}
\end{figure}

We apply the Empirical Mode Decomposition (EMD) algorithm \cite{Chang0605059} to the reconstructions.  EMD is a novel adaptive method for separating a nonlinear time series into components based on instantaneous frequency.  Basically it acts as a dyadic filter bank \cite{Flandrin04}; the set for the best reconstruction case are illustrated in Figs. \ref{fig:EMD}.  We refer to these sets as $EMD_L$ and $EMD_R$. The original data's intrinsic mode functions (IMFs) and residuals (set $EMD_O$) are also shown and for comparison, we present the effect of a simple minded linear interpolation between the endpoints of the known data on the IMFs.

Upon visual inspection, we see that the original data generates 9 components: 8 intrinsic modes and 1 residual (the stopping criterion for our EMD implementation is the same as in Huang {\em{et al}} \cite{Huang98}).  On the other hand, the interpolated data have only 8 components.  Numbering the IMFs from highest instantaneous frequency to lowest, starting from IMF 1, we can see by inspection that both $EMD_L$ and $EMD_R$ are strongly similar to $EMD_O$ in IMFs 4,5,6 and 7.  The differences appear in the higher frequency components, due to the discontinuity between the inserted segment and the unadulterated data.  The endpoint discontinuities evidently modify the variances of IMFs 1,2 and 3, although it seems that they are only affected in the area local to the discontinuity, per IMF.

By way of comparison, a linear interpolant between the edges of the known data show that there is contamination all through the IMFs.  It is so strong that IMF 7, which in the other sets clearly distinguishes the baseline rise and fall of the turbulence over the interval under study, is unable to pick out a clean pedestal.

\section{ARIMA}
\label{sect:ARIMA}
Instead of decomposing the data sequence into KL components, an alternative method studied is that of ARIMA (Auto--Regressive Integrated Moving Average) \cite{Beran94}.  This technique's principle is founded on taking advantage of the past (first moment) behaviour of stochastic data to predict the future characteristics.  It is a well known technique used in econometrix, and we have employed it here to investigate its effects on EMD.

An ARIMA($p,d,q$) process is the solution of the following equation
\begin{equation}
\phi(B) (1-B)^d X_t = \psi(B) \epsilon_t
\end{equation}
where $BX_t = X_{t-1}$ is the backshift operator, $\epsilon_t$ represents a white noise process, and $\phi,\psi$ are polynomials of degree $p$ and $q$ respectively.  For mathematical convenience we take $p$ and $q$ to be zero.  Fractional $d$ generalises the differencing parameter between data points to obtain the long range dependence.  In general, $0 \leq d \leq 0.5$, where longer memory is represented by higher values of $d$.

\subsection{ARIMA and EMD}
\begin{figure}[!htp]
\begin{center}
\begin{tabular}{c}
\includegraphics[height=0.4\textwidth]{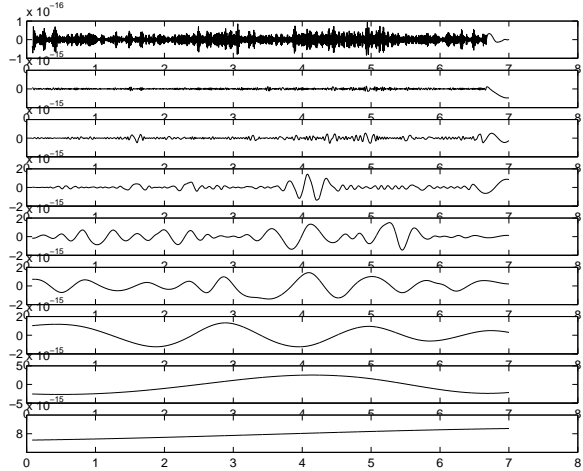}
\end{tabular}
\end{center}
\caption{\label{fig:EMDARIMA} The IMFs and the residual trend line of $EMD_O$, generated by applying Empirical Mode Decomposition to ARIMA(0,$d=0.5$,0).}
\end{figure}
We applied an ARIMA(0,$d$,0) model to predict the behaviour of the final section of the split test data.  The length of each section is 119 points long; by basing the data behaviour on the previous 2373 points and by setting the difference operator to be $d = 0.5$ we generated the IMF results shown in Fig. \ref{fig:EMDARIMA}.

As can be seen, the effect of the ARIMA($0,d,0$) extrapolation is to fill in the higher order IMFs in a smooth fashion (apparently, by visual inspection , IMFs 5 to 8 are smoothly filled at the tail).  It is only untrustworthy when considering the higher frequency components, because ARIMA considers only the conditional first moment.  Note that the number of IMFs and residual are the same as the unadulterated signal $EMD_O$.

\section{CONCLUSIONS}
We have discussed a data hole filling method for stochastic data, a necessary step to be able to use the new technique of Empirical Mode Decomposition upon a time series record.  The Karhunen--Lo\`eve eigenvectors from an ensemble of neighbouring sections of complete data around a data hole can be used to reconstruct the missing segment to a reasonable degree of accuracy, at least for the purposes of applying EMD.  We have shown that the edge continuity is important, although the effect of discontinuities is not universal through all the intrinsic modes of the data.  The quality of the reconstruction is much better using PCA than a simplistic linear interpolant (or merely ignoring the data gap).  We compared this to a simplified ARIMA(0,$d$,0) model, which performed better than the linear interpolant but less effectively than the KL algorithm, disregarding edge effects.  

In summary, when a data hole is present, merely ignoring the data hole or filling it with a linear interpolant is a poor technique from the point of view of EMD.  The result of doing so leads to leakage of spurious effects both laterally in time (per IMF) and longitudinally across all the IMFs.  By reconstructing the datagap with ARIMA($0,d=0.5,0$) one can limit the leakage laterally and longitudinally, although the lowest IMFs remain strongly contaminated with artificial structure.  KL reconstruction limits the leakage best, and even without considering the continuity between the adjacent sections and the reconstructed datagap, it promises to provide the best reconstruction of the three methods described in this paper.


\newpage

\begin{thebibliography}{11}

\bibitem{Santiago05}
F.~Santiago, M.~P.~J.~L. Chang, C.~O. Font, E.~A. Roura, C.~Wilcox, and S.~R.
  Restaino, ``Low altitude horizontal scintillation measurements of atmospheric
  turbulence over the sea: Experimental results,'' {\em Proc. SPIE} {\bf 6014},
  2005.

\bibitem{Font0605050}
C.~O. Font, M.~P. J.~L. Chang, E.~Oh, and G.~C. Gilbreath, ``Humidity
  contribution to the refractive index structure function {$C_n^2$},'' in {\em
  Atmospheric Propagation III},  C.~Y. Young and G.~C. Gilbreath, eds., {\em
  Proc. SPIE} {\bf 6215}, 2006.

\bibitem{Chang0701022}
M.~P. J.~L. Chang, C.~O. Font, G.~C. Gilbreath, and E.~Oh, ``Humidity
  contribution to {$C_n^2$} over a 600m pathlength in a tropical marine
  environment,'' {\em Proc. SPIE} {\bf 6457}, 2007.

\bibitem{Chang06AppOp}
M.~P. J.~L. Chang, C.~O. Font, G.~C. Gilbreath, and E.~Oh, ``Humidity's
  influence on visible region refractive index structure parameter {$C_n^2$},''
  {\em Applied Optics (accepted), ArXiv Physics e-prints} {\bf
  physics/0606075}, June 2006.

\bibitem{Font06MS}
C.~O. Font, ``Understanding the atmospheric turbulence structure parameter
  {$C_n^2$} in the littoral regime,'' Master's thesis, University of Puerto
  Rico at Mayag{\"u}ez, 2006.

\bibitem{Huang98}
N.~E. Huang, Z.~Shen, S.~R. Long, M.~C. Wu, H.~H. Shih, Q.~Zheng, N.-C. Yen,
  C.~C. Tung, and H.~H. Liu, ``The empirical mode decomposition and the
  {Hilbert} spectrum for nonlinear and non-stationary time series analysis,''
  {\em Proc. R. Soc. Lond. Ser. A} {\bf 454}, pp.~903--995, 1998.

\bibitem{Flandrin04}
P. Flandrin, G. Rilling and P. Gon\,calves,
  ``Empirical mode decomposition as a filter bank,''
  {\em Signal Processing Letters, IEEE} {\bf 22}, pp.~112--114, 2004.

\bibitem{Turk91}
M.~Turk and A.~Pentland, ``Eigenfaces for recognition,'' {\em Journal of
  Cognitive Neuroscience} {\bf 3}, pp.~71--86, 1991.

\bibitem{Sirovich87}
L.~Sirovich and M.~Kirby, ``Low--dimensional procedure for the characterization
  of human faces,'' {\em Journal of the Optical Society of America A} {\bf 4},
  pp.~519--524, 1987.

\bibitem{Goodman85}
J.~W. Goodman, {\em Statistical Optics}, Wiley--Interscience, New York, 1996.

\bibitem{Mandelbrot83}
B.~Mandelbrot, {\em The Fractal Geometry of Nature}, W.~H. Freeman and Co., New
  York, 1983.

\bibitem{Roberts96}
A.~J. Roberts and A.~Cronin, ``Unbiased estimation of multi--fractal dimensions
  of finite data sets,'' {\em Physica A} {\bf 233}, pp.~867--878, 1996.

\bibitem{Stanley99}
H.~E. Stanley, L.~A.~N. Amaral, A.~L. Goldberger, S.~Havlin, P.~C. Ivanov, and
  C.-K. Peng, ``Statistical physics and physiology: Monofractal and
  multifractal approaches,'' {\em Physica A} {\bf 270}, pp.~309--324, 1999.

\bibitem{Chang0605059}
M.~P. J.~L. Chang, E.~A. Roura, C.~O. Font, E.~Oh, and C.~Gilbreath, ``Applying
  the {Hilbert-Huang Decomposition} to horizontal light propagation {$C_n^2$}
  data,'' in {\em Advances in Stellar Interferometry, {\em{these
  proceedings}}},  {\em Proc. SPIE} {\bf 6268}, 2006.

\bibitem{Beran94}
J.~Beran, {\em Statistics for Long--Memory Processes}, Chapman and Hall, New
  York, 1994.

\end{thebibliography}
\end{document}